\input harvmac
\pretolerance=10000

\def\B{Bogomol'nyi}
\def\alphadot{\dot \alpha}
\def\betadot{\dot \beta}
\def\phi{\varphi}

\Title{HWS-9720,hep-th/9711154}
{\vbox{\centerline{Fermi-Bose Cancellation in Topologically}
       \vskip2pt\centerline{Non-Trivial Backgrounds}}}

\ifx\answ\bigans{
\centerline{Zvonimir Hlousek\footnote{*}{hlousek@beach.csulb.edu} }
\medskip\centerline{Department of Physics and Astronomy}
\centerline{California State University, Long Beach}
\centerline{Long Beach, CA \ 90840 USA}
\vskip.2in
\centerline{Donald Spector\footnote{$^\dagger$}{spector@hws.edu} }
\medskip\centerline{Department of Physics, Eaton Hall}
\centerline{Hobart and William Smith Colleges}
\centerline{Geneva, NY \ 14456 USA}
}
\else {

\centerline{\qquad\qquad Zvonimir
Hlousek\footnote{*}{hlousek@beach.csulb.edu}
\hfill Donald Spector\footnote{$^\dagger$}
{spector@hws.edu}\qquad\qquad}
\bigskip\centerline{\qquad\qquad Department of Physics and
Astronomy
\hfill Department of Physics, Eaton Hall\qquad\qquad }
\centerline{\qquad\qquad California State University, Long Beach
\hfill Hobart and William Smith Colleges\qquad\qquad }
\centerline{\qquad\qquad Long Beach, CA 90840 USA \hfill Geneva, NY
14456 USA\qquad\qquad }
}
\fi

\vskip .3in
We show in a model-independent way that, in the background of a
topological soliton or instanton that saturates a Bogomol'nyi
bound, the fermion and boson excitation spectra of non-zero modes
cancel at the one-loop level.
This generalizes D'Adda and DiVecchia's result
for some specific instanton models.  Our method also 
establishes, again in a
model-independent way, the generality of the connection between 
zero modes in topologically non-trivial backgrounds and 
and index theorems.

\Date{10/97}

\newsec{Introduction}

Some years ago, D'Adda and DiVecchia showed that there was a 
fermi-bose cancellation when considering
from the excitation spectrum around a self-dual or anti-self-dual 
Yang-Mills instanton field background 
\ref\dd{A. D'Adda and P. DiVecchia, Phys. Lett. 73B (1978) 162.}. 
They found that, except for zero modes, scalars and spinors
in the same
representation of the gauge group had  exactly the right spectra to
cancel at the one loop level, as did spinors and vectors.

Their paper showed this result was true in two particular models, 
but gave no indication
as to whether it was an accident of those particular models or a 
consequence of some more general field theoretic properties. In 
this letter, we will show that the latter possibility obtains, that 
there are general reasons why \B-saturating topologically non-trivial
backgrounds produce such a cancellation.

The occurrence of a cancellation between fermionic and bosonic modes
signals a supersymmetric explanation, and that is exactly what we
present
in this paper.  The reader should note, however, that the explanation
is valid for non-supersymmetric theories with solitons or instantons; it
is only our method that invokes supersymmetry, but, as we will show, the
results are valid quite generically in supersymmetric and
non-supersymmetric
field theories alike.

The essence of our argument is as follows.  
If one wishes to understand the one-loop excitation spectrum in a 
theory, it is sufficient to study any theory that is equivalent 
at the one-loop level. (Working at one-loop means working with 
linearized field equations.)  We will identify a supersymmetric 
theory which agrees with our original theory to one loop. In 
that theory, then, we will be able to use supersymmetry to relate 
the bosonic and fermionic excitations in the soliton or instanton 
background.  Since the theories are identical at one loop, if we can
use supersymmetry to obtain the cancellation in the supersymmetric 
case, we automatically obtain the cancellation in the non-supersymmetric
case.

We will first make an observation or two about the Yang-Mills 
case; then we will take up the general problem, treating solitons 
first and instantons second.  We note too that, to simplify the
terminology, we will sometimes use ``soliton'' and ``instanton''
to mean `` topological soliton'' and ``topological instanton,''
respectively.

\newsec{Yang-Mills and D'Adda-DiVecchia}

Consider $3+1$ dimensional Yang-Mills theory. As is well-known, 
the theory possesses topologically non-trivial field 
configurations known as 
instantons \ref\inst{A.A. Belavin, A.M. Polyakov, A.S. Schwartz, 
and Yu.S. Tyupkin, Phys. Lett B59 (1975) 85\semi
G. 't Hooft, Phys. Rev. Lett. 37 (1976) 8\semi
G. 't Hooft, Phys. Rev D14 (1976) 3432\semi
A.M. Polyakov, Nucl. Phys. B120 (1977) 429.}, and in 
addressing gauge theories non-perturbatively, understanding 
physics in an instanton background is essential 
\ref\reviews{S. Coleman, ``The Uses of Instantons'' in
{\it The Whys of Subnuclear Physics}, A. Zichichi, ed.
(Plenum Press, NY, 1979)\semi
R. Rajaraman, {\it Instantons and Solitons} (North Holland, Amsterdam,
1982).}. It
is straightforward to see that, in an appropriate normalization, 
the Euclidean action is bounded from below by the magnitude of the 
instanton number, and that (anti)self-dual gauge fields saturate 
such a bound. 

In \dd, the authors studied the equations of motion in
the background of a self-dual Yang-Mills instanton.  For scalars $\phi$,
spinors $\psi$, and vectors $A_\mu$, respectively, they found that 
the equations of motion
\eqnn\coveqn
$$\eqalignno{
D_\mu D^\mu \phi &= -\eta_s \phi \cr
i \gamma^\mu D_\mu \psi &= \eta_f \psi \cr
D_\mu F_{\mu\nu} +[F^0_{\nu\mu},A_\mu] &= -\eta_A A}$$
(where $D_\mu$ is the gauge covariant derivative in the instanton
background, and $F^0_{\nu\mu}$ is the field strength of the 
instanton background) give rise to a essentially the
same set of eigenvalues.
In particular, if $\phi$ and $\psi$ are in the same representation
of the gauge group, then for each non-zero eigenvalue $\eta_f$
for the spinor equation, there is a corresponding solution of the
bosonic equation with eigenvalue $\eta_s = \eta_f^2$.
This in turn guarantees that the non-zero modes of the scalar and
the spinor cancel at the one loop level with suitable matter content
(that is, if there are as many scalar as spinor degrees of freedom).

Note that \dd\ also found a corresponding relation between the fermion
and vector non-zero modes, provided the fermion is, like the vector
field, in the adjoint representation.  All the results in \dd\ were
obtained by explicit calculation.

We offer here another way to understand these findings, one 
that is of a more algebraic nature.  This explanation is meant
to motivate the model-independent arguments that appear in the
subsequent sections.

Consider supersymmetric
Yang-Mills theory coupled to matter. In $3+1$ 
dimensional supersymmetry, one has left- and right-handed 
supercharges, $Q_\alpha$ and ${\bar Q}_{\alphadot}$, respectively. Any 
gauge field strength can be decomposed into its self-dual and 
anti-self-dual parts, $F_{\alpha\beta}$ and
$F_{\alphadot\betadot}$. It is a simple calculation to
see that
\eqn\sdsusy{[Q_\alpha,F_{\alpha\beta}]=0 \qquad and 
    \qquad [{\bar Q}_{\alphadot},F_{\alphadot\betadot}]=0~~~.}
Without loss of generality, let us consider the self-dual 
background.

Suppose we have a supersymmetric gauge theory, with gluon $A_\mu$ and 
gluino $\lambda$ both in the adjoint representation, and matter 
fields in some representation ${\tt R}$ and labeled by a scalar $\phi$ 
and a spinor $\psi$. To one-loop order, the equations
of  motion for these various fields are of exactly the same form as
\coveqn, namely
\eqnn\scoveqn
$$\eqalignno{
D_\mu D^\mu \phi -m^2 \phi&= -\eta_s \phi \cr
(i \gamma^\mu D_\mu -m)\psi &= \eta_f \psi \cr
i \gamma^\mu D_\mu\lambda &= \eta_\lambda \lambda \cr
D_\mu F_{\mu\nu} +[F^0_{\nu\mu},A_\mu] &= -\eta_A A~~~.}$$
To this order, only the minimal couplings
to the gauge fields matter; Yukawa couplings and terms 
in the scalar potential other than
mass terms are irrelevant.  Note that this means that in
the massless limit for the matter fields, the linear equations
of motion are the same for all theories with matter fields in
the same representations, regardless of whatever additional
couplings there might be.

Now suppose we have a solution to, say, the scalar equation of
motion, with eigenvalue $\eta_s$.  A finite supersymmetry transformation
with $Q_\alpha$ leaves the background invariant, while
``rotating'' the $\phi$ field into a $\psi$ field.  Since supersymmetry
is an invariance of the theory, the resulting spinor field configuration
will also be a solution to the equations of motion.

What does this mean for the eigenvalues $\eta_f$?  Strictly speaking,
because of the $\gamma^\mu$ in the $\psi$ equation in \coveqn and
\scoveqn, the linearized fermion equation is not an eigenvalue equation,
as the chiralities of the two sides differ.  On the other hand, repeated
iteration {\it does} produce an eigenvalue equation for a single
chirality of $\psi$, so it must
be that $\eta_f^2=\eta_s$ in this particular case.  In other words,
the $\eta_s$ values must be the squares of the $\eta_f$ values
in \coveqn.  (In the presence

A similar relation would hold between the gluino and gluon spectra,
for exactly the same reason.

In this way, we see that the spectra of scalar and spinor
excitations, and of the spinor and vector excitations,
about a self-dual background 
cancel in supersymmetric Yang-Mills
theory. This result by itself is  not terribly surprising.
However, the key observation is this: the  equations for the 
excitation spectrum to one-loop order  are identical for any 
minimally coupled gauge theory, depending  only on the mass and 
the gauge group representation. Matter self-interactions have no
effect on these equations.
So, the bose-fermi  cancellation of the excitation spectrum 
to one-loop order in the 
supersymmetric case (which we can prove by taking advantage of 
supersymmetry) automatically implies the corresponding
cancellation in the non-supersymmetric case.

What is different about the zero modes (which of course arise
only for massless fields)?  A field
configuration of a zero mode plus the self-dual instanton solution
still saturates the \B{} bound.  As a consequence, the supercharge that
leaves the background unchanged also annihilates the zero mode.
Thus there is no way to use supersymmetry to rotate one zero mode into
another.  In fact, it is exactly this reasoning that
enables us to connect the number of zero modes to the index of the
charge $Q_\alpha$, as the zero modes are those non-zero field
configurations in the instanton background that are annihilated
by $Q_\alpha$.
And, since the zero modes of the instanton are
linked to the index of an operator to one-loop order in the
supersymmetric theory, they are linked to the index of that operator to
one-loop order in the large range of non-supersymmetric theories that
have the same one-loop field equations.

The above results are, in the case of Yang-Mills theory, well-known.
However, our explanation, through the use of the superalgebra and
in a way that connects the non-supersymmetric cases to the
supersymmetric results, lays the groundwork for a generic approach
to this question.  Indeed, the same structure presented here in
this case appears quite generically in a range of soliton and
instanton models.  In the remainder of this paper, we obtain
just such a model-independent argument for these features of
the excitation spectra in \B-saturating backgrounds.

\newsec{General Argument: Solitons}

While we have discussed the Yang-Mills instanton above, as it is
perhaps the most familiar case, for the purposes of our general
argument we will begin first with soliton backgrounds, and then
generalize our results to instanton backgrounds.

Our goal is to understand in a model-independent way the cancellation
between bosonic and fermionic excitations
about \B-saturating solitons at the
one-loop level.  Our approach will be as follows.
Suppose we have
a theory that has non-trivial soliton field configurations among
its solutions.  We wish to consider the one-loop boson and
fermion excitation spectra.  Our plan will be to link this theory to
a supersymmetric theory, and then to use algebraic methods in the
supersymmetric theory to understand the excitation spectrum.  Since
the theories will have the same one-loop physics, the results for the
excitation spectrum in the supersymmetric theory will automatically
imply the identical results for the original (non-supersymmetric) 
theory.

For simplicity, for concreteness we will refer at points to a scalar
$\phi$ and a spinor $\psi$, but any fields that can be superpartners
would work for the discussion that ensues here.

So suppose we have a generic theory with topologically non-trivial
soliton field configurations among its solutions.
In \ref\bogex{Z. Hlousek and D. Spector,
Nucl. Phys. B397 (1993) 173.} and \ref\sinew{Z. Hlousek and D. Spector, 
Nucl. Phys. B442 (1995) 413.}, it is
established that topological solitons in three or more spacetime
dimensions will exhibit \B{} bounds.  Thus we will take it that
we are in such a situation, in the presence of a \B-saturating
soliton.

We now wish to link this theory to a supersymmetric theory.  In
\sinew, it is shown that there is necessarily a supersymmetric
extension to a theory with solitons.  However, we need a slightly 
different result here.  Let us
truncate the theory we are studying to the one-loop level.  This
theory will have solitons, still, since a topological charge
is conserved without reference to the equations of motion, and will
also have fields $\phi$ and $\psi$.  We now require that there be a
supersymmetric extension of this truncated theory in which $\phi$ and
$\psi$ are superpartners.  Of course, this means, for example, that
these fields must have the same (possibly zero) mass.  We will refer
to this supersymmetric theory as the 
associated supersymmetric theory of
the original theory.\foot{Because of the large variety of possible
superfield representations, especially when considering different
dimensions, it is not feasible to give a generic construction of the 
associated supersymmetric theory.  However, in all typical cases with
which we are familiar, such as Yang-Mills theory and sigma models, it
is quite simple and straightforward to find the associated
supersymmetric theory, with no obstacles to
this construction.  Thus the existence of associated supersymmetric
theory seems something one can rely on.}

Our goal then is to show that the bose-fermi cancellation occurs in
the associated supersymmetric theory, and then that this occurs as
well in the original theory.

The study of topological solitons in supersymmetric theories
was pushed forward by \ref\ow{D. Olive and E. 
Witten, Phys. Lett. B 78 (1978) 97.}, which identified
some theories in which solitions appeared to be associated with
extended superalgebras.  This phenomenon was ultimately explained
in a general setting, independent of particular models, 
in \ref\n2susy{Z. Hlousek and D. Spector, Nucl. Phys. B370 (1992) 143.}
and \sinew.  These papers demonstrated that any theory
with  $N=1$ supersymmetry and a topologically conserved charge in
$2+1$  or more dimensions automatically is endowed with a richer
supersymmetry invariance,  namely invariance under centrally
extended $N=2$ supersymmetry in which the  topological charge is the
central charge.  This result followed from considering the
superfield that includes the potential for the conserved
topological current \ref\topcurrent{C. Cronstr\"om 
and J. Mickelsson, J. Math. Phys. 24 (1983) 2528\semi
R. Jackiw, in {\it Current Algebra and Anomalies}, 
S. Treiman, R. Jackiw, B. Zumino, and E. Witten, ed. 
(World Scientific, Singapore, 1985)}.

The consequences of this extended superalgebra are significant. On 
algebraic grounds alone, one can show that the energy of such 
solitons is bounded from below by their topological charge (in an 
appropriate normalization which the algebra determines), and that 
the solitons which saturate such a bound are annihilated by one 
linear combination of the supersymmetries. Thus the states in the 
theory for the most part come in conventional $N=2$ multiplets; 
but the states that saturate this Bogomol'nyi-type bound appear 
in reduced multiplets, which look like $N=1$ multiplets and which,
most importantly for our purposes,  are annihilated by half the
superalgebra 
(see \ref\n2alg{A. Salam and J. Strathdee, Nucl. Phys. B80
(1974) 499.}).

So imagine that we have constructed the associated supersymmetric
theory. It is perhaps worth noting that since the
topological current of the original theory is conserved without
reference to the equations of motion (this is what is meant by a
topological current), it remains a conserved current in the associated
supersymmetric theory; the agreement of the theories to the one-loop
level also means that, to the necessary accuracy, the energy of the
soliton is the same in both theories.

We can now obtain the fermi-bose cancellation in the solitonic
background in the associated supersymmetric theory.  Suppose we
have a \B-saturating soliton in this theory. We now consider an
excitation of the scalar field $\phi$ that satisfies the one-loop
equations of motion with eigenvalue $\eta_s$.  We know 
from the $N=2$ superalgebra that there is a supercharge that
annihilates the \B-saturating soliton; let us call this supercharge
$Q$ (suppressing a spinor index).  Consider applying $Q$ to the field
configuration consisting of the soliton and the scalar excitation. 
Since $Q$ annihilates the soliton but maps $\phi$ into $\psi$, if
we consider a finite supertransformation generated by $Q$, we will
leave the soliton background unchanged, and, since the equations of
motion are invariant under the superalgebra of the theory, we will
therefore map a scalar solution with this particular soliton
background into a fermionic solution for the exact same background.

Thus the bosonic and fermionic solutions are paired.  Since the
one-loop equations are the same as in the original theory, the solutions
are paired in that original theory, too, even though it has not
supersymmetric.

Let us return to the associated supersymmetric theory and consider
the eigenvalues.  Because fermions are represented by spinors, the
fermion equation of motion typically needs to be iterated twice to
get the actual physical eigenvalues.  (This phenomenon is not special
to our situation; note that the mass term for a fermion has one power
of the mass, while that for a boson has two.)  Thus the equation of
motion obtained through the variation $\delta S/\delta\psi$ will
yield an equation of the form $\hat\Omega \psi = \eta_f\psi$
that will mix fermionic components; iterating $\hat\Omega$ till we
obtain an equation that does not mix components---which will
yield the relation that $\eta_f^2=\eta_s$.  

Alternatively, if one wishes to think in a functional language, 
consider the determinants that arise in the one-loop functional
integral.  These are the  determinants
$\delta^2 S/\delta\psi^2$ and $\delta^2 S/\delta\phi^2$, where
$S$ is the action of the associated supersymmetric theory.
Since $S$ is invariant under supersymmetry transformations, since
the soliton background is also invariant under a supersymmetry
transformation, and since that supersymmetry transformation
pairs $\phi$ with $\psi$, these determinants will
have to cancel in the functional integral.  This is precisely
what the condition $\eta_f^2=\eta_s$ ensures.

Thus in the associated supersymmetric theory, supersymmetry has
guaranteed that the bosonic and fermionic excitations will cancel at
the one-loop level. Since this theory and the original theory have
the same one-loop equations of motion, this means that the bose-fermi
cancellation occurs in the original non-supersymmetric theory as well.

How are the zero modes different?  These modes still saturate
the \B{} bound, and so they are annihilated by the same supercharge
that annihilates the soliton background.  Hence we cannot rotate
these solutions into new superpartner solutions to the equations of
motion.  This also tells us, then, that the zero modes in the soliton
background arise from the kernel of the supercharge $Q$ that
annihilates the soliton background in the associated supersymmetric
theory, and so the zero modes are related to the index of this
operator.  Since the zero modes in this soliton background are the same
at the one-loop level in the original and the associated supersymmetric
theory, this means that even in the original theory, the zero modes are
associated with the index of the operator $Q$.  This explains the
regular appearance of index theorems in understanding the zero modes in
topologically non-trivial soliton backgrounds. Of course, the original
theory has no conserved supercharge;
but  the natural appearance of an index theorem interpretation of the
zero modes in the soliton background demonstrates the intimate
connection between the original theory and
its supersymmetric associate.

\newsec{General Argument: Instantons}

The generalization of these results to instantons in a
model-independent way requires
connecting the instanton results to soliton results, rather than
directly making a connection to supersymmetric theories with
instantons, despite our preliminary discussion of the Yang-Mills case. 
The reason is straightforward.  Solitons are associated with a
conserved charge, and this conserved charge, as we have discussed, gets
integrated into the supersymmetry algebra of the associated 
supersymmetric theory.  Consequently, what was
previously an argument about equations of motion
(model dependent) becomes an argument about algebras and their
representations (model independent).  Instantons, on the other hand,
though they fit into a topological classification, do not correspond to
a conserved charge, and so there is no possibility that instanton
number will become part of an algebra.

Instead, we follow the guidance of \sinew, and link the instanton
case to the soliton case.  

So suppose we have a theory with instantons. 
This means that we have some Euclidean action in $d$ dimensions that
possesses a topological classification of field configurations.

Let us imagine that we add a time dimension, turning this into a
$d+1$-dimensional Minkowski theory.  Of course, this additional
dimension may well necessitate additional fields, too, to maintain
Lorentz invariance: e.g., vector fields will need an additional
component, and fermion representations might need to be enhanced.

Because the original instanton classification in $d$ dimensions was
topological, this classification persists in the Minkowski theory,
but, because of the additional time dimension, it corresponds to a
conserved topological charge in $d+1$ dimensions. (At each time
slice, the instanton classification can be applied, and continuity
ensures that time evolution cannot switch the system from one sector
to another of this topological classification.)

Thus, the enhanced Minkowski theory has topological solitons.  Among
these solitons, the static solitons play a special role: these are
nothing but the instanton configurations of the original 
theory.\foot{To be precise, we single out the soliton
configurations that involve only the field components of the original
theory, and not the extra components that might have been added.
We take this extra detail to be understood.}

Let us now consider the enhanced theory.  We know from the preceding
section that the bosonic and fermionic excitations in a
\B-saturating soliton background are paired and hence cancel each
other.  We can refine this pairing a bit further.

Let us consider a static soliton background of the enhanced theory
that saturates the \B{} bound.  This field configuration is
automatically an instanton of the original theory, and one that
automatically saturates the corresponding \B{} action bound of the
original Euclidean theory.  (See \sinew\ for a more detailed
discussion of this connection.)  Now consider, in the 
dimensionally enhanced theory,
the time-independent excitations about this soliton background.  
The excitations in this subclass 
are paired and hence cancel among each other at the one-loop
level.  The
reason is that if one constructs the associated supersymmetric theory
for this already dimensionally enhanced theory, in the associated
supersymmetric theory, supersymmetry maps time-independent excitations
into other time-independent excitations.  (Recall that supersymmetry
commutes with the Hamiltonian, which is the generator of time
translations.)  Thus not only are the bose and fermi excitations paired
as we saw in the previous section, but in fact
the time-independent excitations are 
paired and therefore cancel among themselves.

But this is just what we need.  The time-independent excitations
about static solitons are paired in the enhanced $d+1$-dimensional
Minkowski theory; mathematically, these are identical to the
excitations about the instantons in the $d$-dimensional theory, and
so these excitations too exhibit a bose-fermi pairing.
This in turn means that the one-loop contribution
between bosons and fermions in a \B-saturating instanton background
will cancel, except for the zero modes.

Thus the original result of
D'Adda and DiVecchia's has finally been obtained in a model
independent way.  In the specific case of Yang-Mills instantons,
this means enhancing the $4$-dimensional Euclidean theory to a
$4+1$-dimensional Minkowski theory, extending this to its 
associated supersymmetric theory, obtaining the pairing between static
excitations in this theory, and then working back down the chain to
the original Euclidean theory.  The requirement that, in
the associated supersymmetric theory of the dimensionally
enhanced theory, the
topological charge appears as the central charge of an extended
superalgebra means that the generic result we have found pairing
bosonic and fermionic non-zero modes in \B-saturated instanton
backgrounds will
hold in two or more Euclidean dimensions.

Note, too, that if we consider the static excitations in the
$d+1$-dimensional theory about the static soliton background, the
static zero modes are annihilated by the $d+1$-dimensional
supercharge.  Since these form the subset of zero modes that are
invariant under time translations (and time translations commute
with supersymmetry), we can identify these zero modes by studying
the index of the relevant supercharge but only in
the subspace of time translation invariant field configurations.
This then associates zero modes in a \B-saturating instanton
background with index calculations through a model-independent
argument, and thus explains in a general way the connection between
such zero modes and index theorems.

\newsec{Conclusions}

We have seen that the phenomenon of bose-fermi cancellation in a
\B-saturating background can be understood in a model independent
way.  Our argument applies to soliton backgrounds in $2+1$ or more
dimensions, and to instanton backgrounds in two or more Euclidean
dimensions. The model independence stems from our ability to use
algebraic arguments to analyze the solutions to the equations of
motion.  We are able to understand non-supersymmetric theories by
means of a certain superalgebra because the non-supersymmetric and
supersymmetric theories produce the same one-loop equations of
motion.  This is very much like the use of complex analysis to
understand properties of real functions, or the use of
supersymmetric Yang-Mills or string theory to calculate ordinary
gluon amplitudes.  We note too that our methods automatically shed
light on the generic appearance of index theorems in
understanding the zero modes in such backgrounds.

It is worth noting that in \dd, the authors link the excitations of
scalars, spinors, and vectors in a Yang-Mills instanton background
in four Euclidean dimensions.  Our arguments explain this extended
degeneracy quite easily.  Suppose we follow our chain of argument,
enhancing the four Euclidean dimensions to five Minkowski
dimensions, adding supersymmetry, and ultimately reducing to four
dimensions again.  If we reduce immediately from the supersymmetric
$4+1$ dimensional Minkowski theory to four dimensions (but keep the
extra field components that had to be added to maintain $4+1$
dimensional Lorentz invariance), we
automatically wind up with a four dimensional theory with explicit
$N=2$ supersymmetry.  (This $N=2$ symmetry comes entirely from the
explicit $N=1$ supersymmetry in five dimensions, not from the
extended superalgebra that the topological charge produced in that
case.)  This $N=2$ supersymmetry in four dimensions allows scalars,
spinors, and vectors in the adjoint representation to be grouped into
one hypermultiplet.  At the linearized level, then, these
scalar, spinor, and vector excitations are grouped, can be rotated
into each other, and thus have a common set of non-zero eigenvalues
in a \B-saturating background.

This research was supported in part by NSF Grant. No. PHY-9509991.

\listrefs
\bye